\documentclass[%
 aip,
 sd,%
 reprint,%
]{revtex4-1}

\usepackage{graphicx}% Include figure files
\usepackage{dcolumn}% Align table columns on decimal point
\usepackage{bm}% bold math
\usepackage[cmex10]{amsmath}
\usepackage[utf8]{inputenc}
\usepackage{float}
\usepackage{amssymb}
\usepackage[T1]{fontenc}
\usepackage{natbib,hyperref}
\usepackage{color,soul}
\usepackage{mathtools}
\usepackage{amsmath}
\usepackage{epsfig}
\usepackage{color}

\begin{document}

\preprint{AIP/123-QED}

\title[Biaxial magnetic-field setup for angular magnetic measurements of  thin films and  spintronic nanodevices]{Biaxial magnetic-field setup for angular magnetic measurements of  thin films and  spintronic nanodevices}

\author{Piotr Rzeszut}
\email{piotrva@agh.edu.pl}
\affiliation{ 
AGH University of Science and Technology, Department of Electronics, Al. Mickiewicza 30, 30-059 Krak\'{o}w, Poland}

\author{Witold Skowroński}
\affiliation{ 
AGH University of Science and Technology, Department of Electronics, Al. Mickiewicza 30, 30-059 Krak\'{o}w, Poland}

\author{Sławomir Zi\ifmmode \mbox{\k{e}}\else \k{e}\fi{}tek}
\affiliation{ 
AGH University of Science and Technology, Department of Electronics, Al. Mickiewicza 30, 30-059 Krak\'{o}w, Poland}

\author{Piotr Ogrodnik}
\email{piotrogr@if.pw.edu.pl}
\affiliation{University of Michigan, Department of Electrical Engineering and Computer Science, Ann Arbor, MI 48109, USA}
\altaffiliation[permanent address: ]{Warsaw University of Technology, Faculty of Physics, , ul. Koszykowa 75, 00-662 Warszawa, Poland}

\author{Tomasz Stobiecki}
\affiliation{ 
AGH University of Science and Technology, Department of Electronics, Al. Mickiewicza 30, 30-059 Krak\'{o}w, Poland}
\affiliation{AGH University of Science and Technology, Faculty of Physics and Applied Computer Science, Al. Mickiewicza 30, 30-059 Krak\'{o}w, Poland}

\date{\today}

\begin{abstract}
The biaxial magnetic-field setup for angular magnetic measurements of thin film and spintronic devices is designed and presented. The setup allows for application of the in-plane magnetic field using a quadrupole electromagnet, controlled by power supply units and integrated with an electromagnet biaxial magnetic field sensor. In addition, the probe station is equipped with a microwave circuitry, which enables angle-resolved spin torque oscillation measurements. The angular dependencies of magnetoresistance and spin diode effect in a giant magnetoresistance strip are shown as an operational verification of the experimental setup. We adapted an analytical macrospin model to reproduce both the resistance and spin-diode angular dependency measurements. 
%
%Valid PACS numbers may be entered using the \verb+\pacs{#1}+ command.
\end{abstract}

\keywords{quadrupole electromagnet, magnetometry of thin magnetic films and nanodevices, magnetoresistance, spin diode effect}%Use showkeys class option if keyword
\maketitle

\section{\label{sec:Introduction}Introduction}
Ferromagnetic resonance (FMR)\cite{kittel1947interpretation} in thin ferromagnetic films measured typically in a microwave regime can deliver much useful information important for spintronics applications\cite{chumak2015magnon} such as magnetization saturation, magnetic anisotropy constant and Gilbert damping. In patterned devices of micro- to nanometer sizes, FMR can be detected electrically using the spin-torque diode (SD) effect.\cite{tulapurkar2005spin,harder2016electrical} In this effect, a radio frequency (RF) current passes through an anisotropic magnetoresistance\cite{yamaguchi2007rectification,ziketek2016electric}, giant magnetoresistance\cite{zietek2015rectification,zietek2015influence,kleinlein2014using} or tunnel magnetoresistance\cite{tulapurkar2005spin,skowronski2014spin} device, which, due to spin transfer or field torque effect, induces the resistance oscillations. These oscillations mix with RF current and as a result produce an output DC voltage.\cite{zietek2015interlayer,tulapurkar2005spin}

The SD effect can potentially be used in various applications, such as 
microvave detectors\cite{skowronski2014spin,wang2009sensitivity,fan2009magnetic}, modulators\cite{pufall2005frequency} and demodulators\cite{yamaguchi2007self}.

In addition, by performing the analysis of angular dependence of SD measurements, one can determine the dynamic properties of the investigated magnetic system, such as resonance fields or frequencies of both FMR\cite{ziketek2016electric} and standing spin wave modes\cite{zietek2017electric}.

An experimental approach often requires a characterization of prototype devices by applying an external magnetic field at different angles in plane or to measure angular relations of electric and magnetic parameters (e.g. resistance vs. magnetic field angle)\cite{zietek2015rectification}. Such measurements can be done by mechanically rotating a sample in a dipole electromagnet, using perpendicular crossed Helmholtz coil pairs or using a quadrupole electromagnet.

The first method requires a complex mechanical system, that is difficult to realize for microwave measurements. 
In perpendicularly crossed Helmholtz coils an arbitrary magnetic field angle can be easily applied by setting suitable currents for each pair of coils. However the magnitude of the field in such non-superconducting electromagnet is limited to a few kA/m.
A quadrupole electromagnet enables the increase of the maximum field magnitude, while still allowing to control the field angle electrically. However, due to the non-linear magnetizing curve of the ferromagnetic cores and strong influence of coil pairs on each other, a precise magnetic-field control system is needed.

In this work, we present a complete microwave and magnetoresistance measurement system that allows for varied angle-resolved determination of spintronic device parameters. Application of a magnetic field up to 160 kA/m is possible at any angle. The field may be controlled using a proportional–integral–derivative (PID) algorithm or without a feedback loop. The measurement setup consists of the following: RF generator, RF spectrum analyser, bias tee, sourcemeter, and lock-in voltmeter. In addition in this work we present a specially designed electromagnet-pole-shape which can increase field homogeneity inside and extend the area where field is homogeneous, without increasing the outer dimensions of the magnet. Dedicated software enables measurement using the presented hardware, especially  DC voltage vs. RF current characterisation.

In Sec. \ref{sec:field_gen_measm} we describe the generation and measurements of magnetic field  by a quadrupole electromagnet and electronics implementation of a magnetic field vector controller. In Sec. \ref{sec:spin_meas_setup} we present the measurement setup used to examine SD voltage and in Sec. \ref{sec:sample} a sample structure. Later we discuss the results of using the designed system and present measurement of SD voltage.

\section{\label{sec:experimental}Experimental}

\subsection{\label{sec:field_gen_measm}Magnetic field generation and measurement}

\subsubsection{\label{sec:quadElectromagnet}Quadrupole electromagnet}
\begin{figure}[h!]
\begin{center}
	\includegraphics[width=\columnwidth]{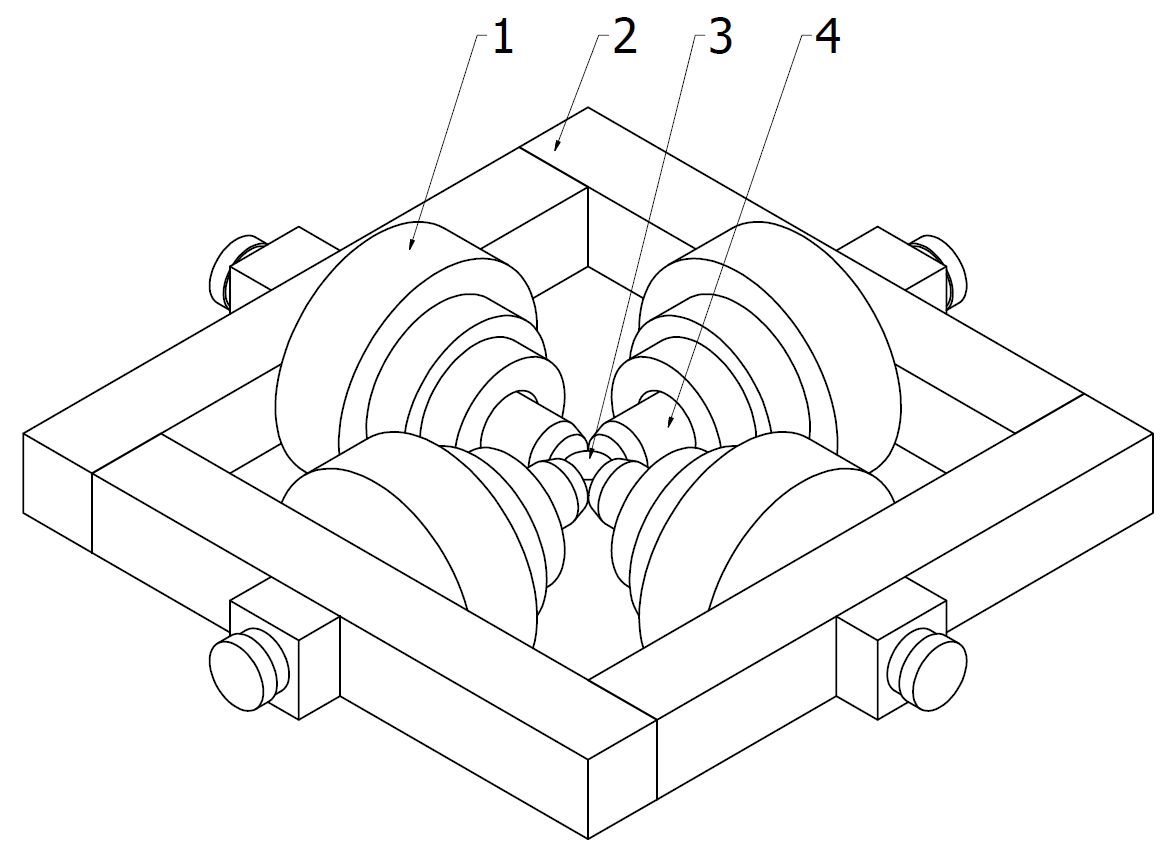}
	\caption{Design of the quadrupole magnet. 1 - coil, 2 - outer frame, 3 - sample holder with bi-axial magnetic field sensor, 4 - cylindrical core with spherical-shaped pole.}
\label{fig:f0}
	\end{center}
\end{figure}

To apply an arbitrary in-plane magnetic field vector, a quadrupole electromagnet is used. Fig. \ref{fig:f0}. presents a drawing of the magnet consisting of: two pairs of coils placed orthogonally to each other and four ferromagnetic cores mechanically connected to the outer frame. Two voltage-controlled current sources (VCCS) are used for application of current in order to generate the required magnetic field vector in the area between the poles.

A key issue in biaxial magnetic field sources is the homogeneity of the field, because it is not always possible to place the element under measurement in the geometrical center between the poles.

Therefore, finite element simulations were used to determine the region suitable for spintronic measurements, taking field homogeneity under consideration. Thin film samples inside the magnet can be treated as two-dimensional objects placed in the geometrical center between the poles, therefore the problem of homogeneity measure can be simplified as two-dimensional. Additionally, as will be mentioned afterwards, the magnetic field vector ($\hat{H}$) in the center of the magnet can be measured and controlled precisely, so $\hat{H}(0,0,\theta_{set})$ can be used as a reference value.

The representation of the field vector in polar coordinates (Eq. \ref{eq:polar}) enables calculation of the error of the magnetic field angle and magnitude, where $x$ and $y$ represent the position in the electromagnet, $\theta_{set}$ expected angle and $\theta_{H}(x,y,\theta_{set})$ actual angle at $(x,y)$ position.

\begin{equation}
\label{eq:polar}
\begin{multlined}
\hat{H}(x,y,\theta_{set})=\\
=\hat{x}|H(x,y,\theta_{set})| \sin \theta_{H}(x,y,\theta_{set}) +\\
+ \hat{y}|H(x,y,\theta_{set})| \cos \theta_{H}(x,y,\theta_{set})
\end{multlined}
\end{equation}

Equations \ref{eq:deltaB} and \ref{eq:deltaTheta} represent the measure of homogeneity using a normalised maximum difference in magnetic field magnitude (or angle) in the geometrical center between the poles and field at the $(x,y)$ position inside the circle of radius $r$.

\begin{equation}
\label{eq:deltaB}
\Delta |H|(r) = \frac{\max_{(x^2+y^2)\leqslant r^2}\big||H(0,0,\theta_{set})|-|H(x,y,\theta_{set})|\big|} {|H(0,0,\theta_{set})|}
\end{equation}

\begin{equation}
\label{eq:deltaTheta}
\Delta \theta_{H}(r) = \frac{\max_{(x^2+y^2)\leqslant r^2}|\theta_{H}(0,0,\theta_{set})-\theta_{H}(x,y,\theta_{set})|} {\pi}
\end{equation}

\begin{figure}[h!]
\begin{center}
	\includegraphics[width=5cm]{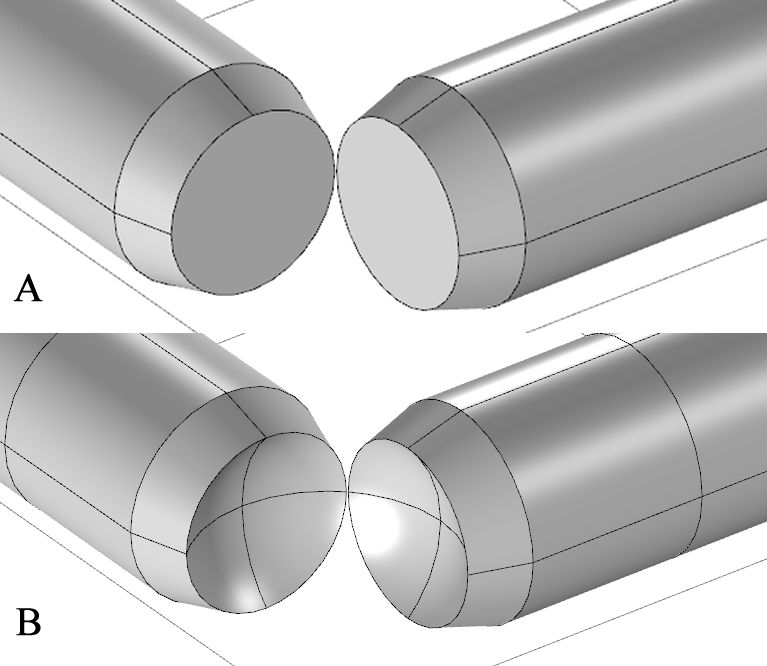}
	\caption{Overview of the poles. A - flat, B - with spherical holes.}
\label{fig:poles}
	\end{center}
\end{figure}

Initially cores with flat poles were prepared, but the achieved homogeneity was not satisfactory. After finite element simulations, the poles with spherical hollows were selected as the best option to improve homogeneity. Both types of poles are presented in Fig. \ref{fig:poles}.

\subsubsection{\label{sec:control_algorithm}Magnetic field control algorithm}
The magnetic field control system contains two VCCS, connected to two independent pairs of coils.
In the center between the poles a bi-axial magnetic field sensor is used for magnetic field vector measurements.

For two perpendicularly crossed pairs of Helmholtz coils, the magnetic field could be controlled by applying current according to simple trigonometric formulas, however, for quadrupole electromagnet these relations are not true, as two pairs of cores mounted in the electromagnet influence the magnetic field produced by each other. Therefore, the angle and magnitude of the field do not follow the rules described above.
Moreover, a step response of the electromagnet was measured using a magnetometer in order to determine the time constant of the electromagnet, which is equal to 600 ms. This relatively high value of the time constant adds complexity to controlling the magnetic field.
The solution for these issues is to use a feedback loop to control the field in the area between the poles. 
For this purpose, two independent discrete PID controllers with anti-windup were used.
\subsubsection{\label{sec:field_controller}Magnetic field controller}
\begin{figure*}
\begin{center}
	\includegraphics[width=\textwidth]{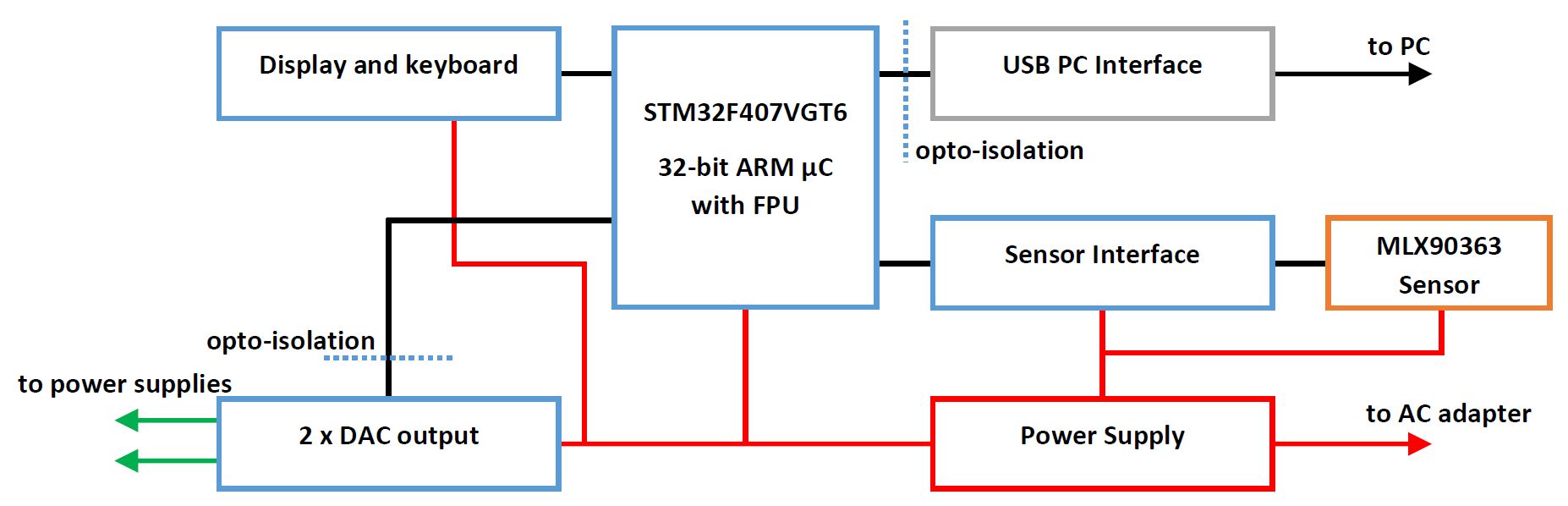}
	\caption{Field control and measurement device overview.}
\label{fig:block_diagram}
	\end{center}
\end{figure*}

The overview of the control and measurement device is presented on Fig. \ref{fig:block_diagram}.

The hall sensor MLX90363 with SPI interface was used as a 3-axial magnetic field sensor. This sensor provides data via a digital SPI interface. It allows for a sampling rate of 250 times per second with 14-bit resolution (13 bits + sign) when filtering and automatic gain control is switched off. The measurement magnetic field range of this sensor is approx. $ \pm$35.8 kA/m for each axis. The sensor is placed on a separate printed circuit board utilizing a cable connection to the device. The board is suitable to be installed inside a sample holder. In this application we use two of the three available axes.  Although the magnetic fields used in the setup exceed the range of the sensor, a simple linear relation of the magnetic field vs. current is used for higher magnetic field values. This approach is justified in the region above the magnetic remamnence and below the saturation of the electromagnet core.
\subsubsection{\label{sec:signal_processing}Control and signal processing}
The ARM microcontroller ($\mu$C) STM32F407 was used as the main processing unit. It has an embedded Floating-Point Unit (FPU), which enables performing high-speed mathematical calculations and is necessary to calculate the field parameters and run two independent PID controllers with sufficient speed.

The communication interface is based on an FT232RL USB to UART converter. The chip provides serial communication (Virtual Com Port) between the PC and the ARM microcnotroller. The interface is powered via USB and optoisolated from the rest of the hardware in order to ensure that the PC will not be damaged, even when some power issues occur. The interface was also used to upgrade the firmware.

\subsubsection{\label{sec:analog_out}Analog output}
Analog driving output signals are generated using DAC8531 (16 bit Digital to Analog Converter with serial interface). The reference voltage is provided by REF193 which outputs 3 V. Using operational amplifiers (OP07CR) -1.5 V voltage is generated, and then added to the output of each DAC. It results in shifting the range of voltage from 0$\sim$3 V to $ \pm$1.5 V. Then a second set of amplifiers uses a gain of $\frac{22}{3}$ to provide output in the range of $ \pm$11 V. Digital control signals are also optoisolated from control logic and powered from a separate voltage regulator to minimize the noise generated by the microcontroller. 

The circuit provides output voltage resolution of approx. 0.34 mV. Considering noise, a real resolution of 1 mV was obtained. Two VCCS are provided by two Kepco power supplies BOP 50-8M. In this configuration a field resolution of 40 A/m was obtained.

\subsubsection{\label{sec:power_supply}Power supply for magnetic field controller}
The power supply is located on an external printed circuit board to allow easy replacement. To avoid introducing high voltages in the system, the controller is supplied by a 5 V and 2 A power supply. Then two AM1P-0512SZ DC converters were used do provide $ \pm$12 V for the analogue output part. Also a low-dropout regulator LM1117-3.3 was used to provide 3.3 V for the microcontroller.

\subsubsection{\label{sec:user_int}User interface}
The display and control module was also made as a separate printed circuit board to allow mounting on a front panel. A HD44780 compatible 20$\times$4 character display was used. The display presents measurement results in each axis, error messages and mathematical functions of field vector: projections to various planes, total magnitude, angles.

\subsubsection{\label{sec:software}Software}
As the measurement process is controlled by the PC, the quadrupole system is also integrated with software designed using the LabVIEW environment. The software, together with the necessary equipment, enables taking the following measurements:

\begin{itemize}
\item $R(H)|_{\theta_{H}=const.}$ - field resolved magnetoresistance
\item $R(\theta_{H})|_{H=const.}$ - angle resolved magnetoresistance
\item $V_{DC}(H)|_{\theta_{H}=const., f=const.}$ - field resolved spin-torque diode
\item $V_{DC}(\theta_{H})|_{H=const., f=const}$ - angle resolved spin-torque diode
\item $V_{DC}(f)|_{H=const., \theta_{H}=const}$ - frequency resolved spin-torque diode
\end{itemize}

The software also supports parametric sweeps allowing to prepare a set of measurements. This is very useful when preparing for example a set of linear magnetic field sweeps with different angles. A graphical interface is provided to select the equipment used and control its parameters.

\subsection{\label{sec:spin_meas_setup}Spin diode measurement setup}

\begin{figure}[h!]
\begin{center}
	\includegraphics[width=\columnwidth]{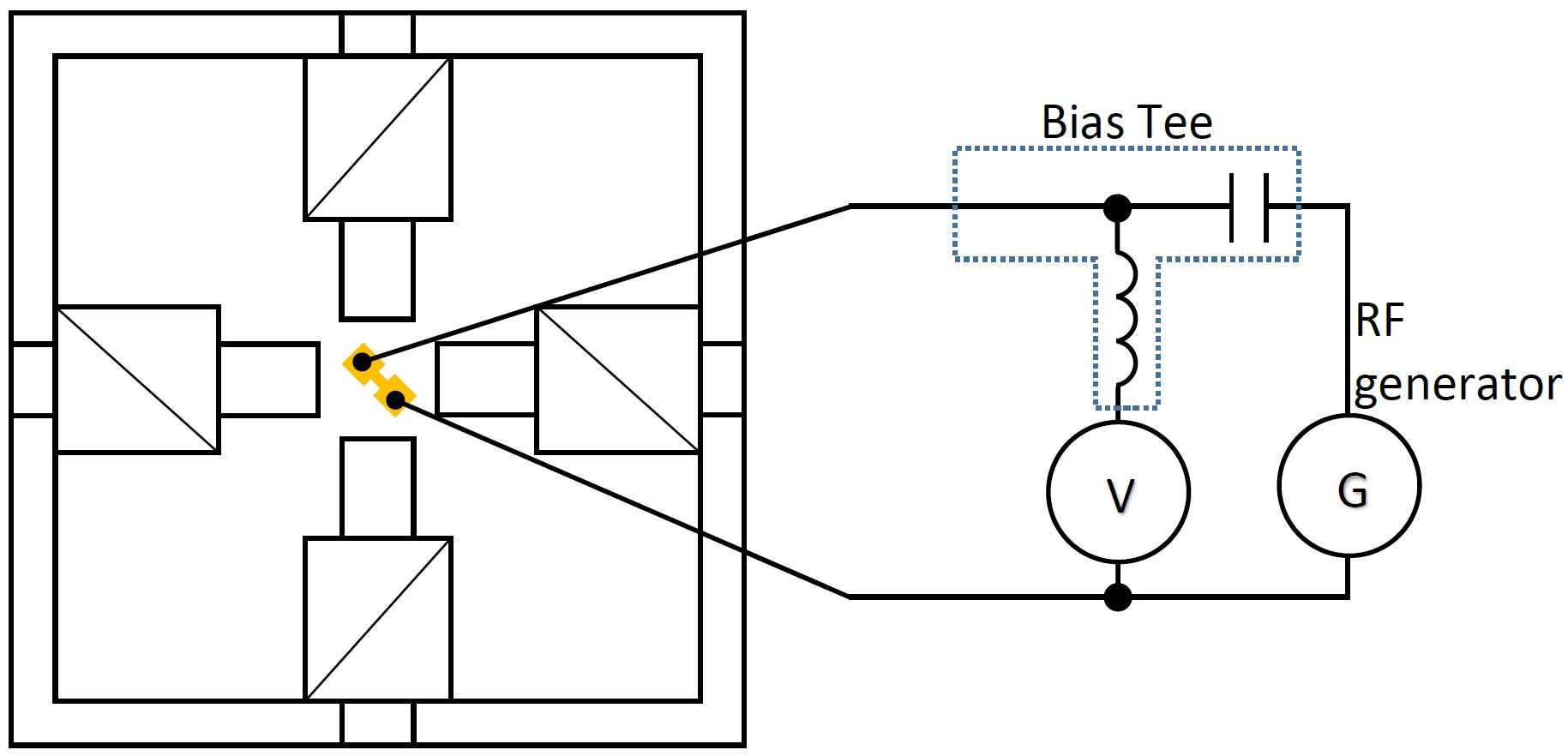}
	\caption{Setup for angular SD effect measurement.}
\label{fig:STFMR_setup}
	\end{center}
\end{figure}

The example experimental setup presented in Fig. \ref{fig:STFMR_setup}. enables angular depedence measurements of both magnetoresistance and SD voltage\cite{zietek2015rectification}. The bias Tee separates the RF input signal from the output DC voltage generated by the examined spintronics device. Also, by replacing the voltmeter with a sourcemeter, simultaneous resistance measurements can be performed. 

The setup utilizes an Agilent RF generator Model PSG E8257D with output power set to 10 dBm, while for DC voltage measurements - an Agilent 34401A voltmeter. The sample is connected to the system using a signal-ground RF probe provided by Picoprobe with $200$  $\mu$m pitch. 20 GHz bandwidth RF cables are used to connect the probe, Bias Tee and generator.

\subsection{\label{sec:sample}Sample}
\begin{figure}[h!]
\begin{center}
	\includegraphics[width=\columnwidth]{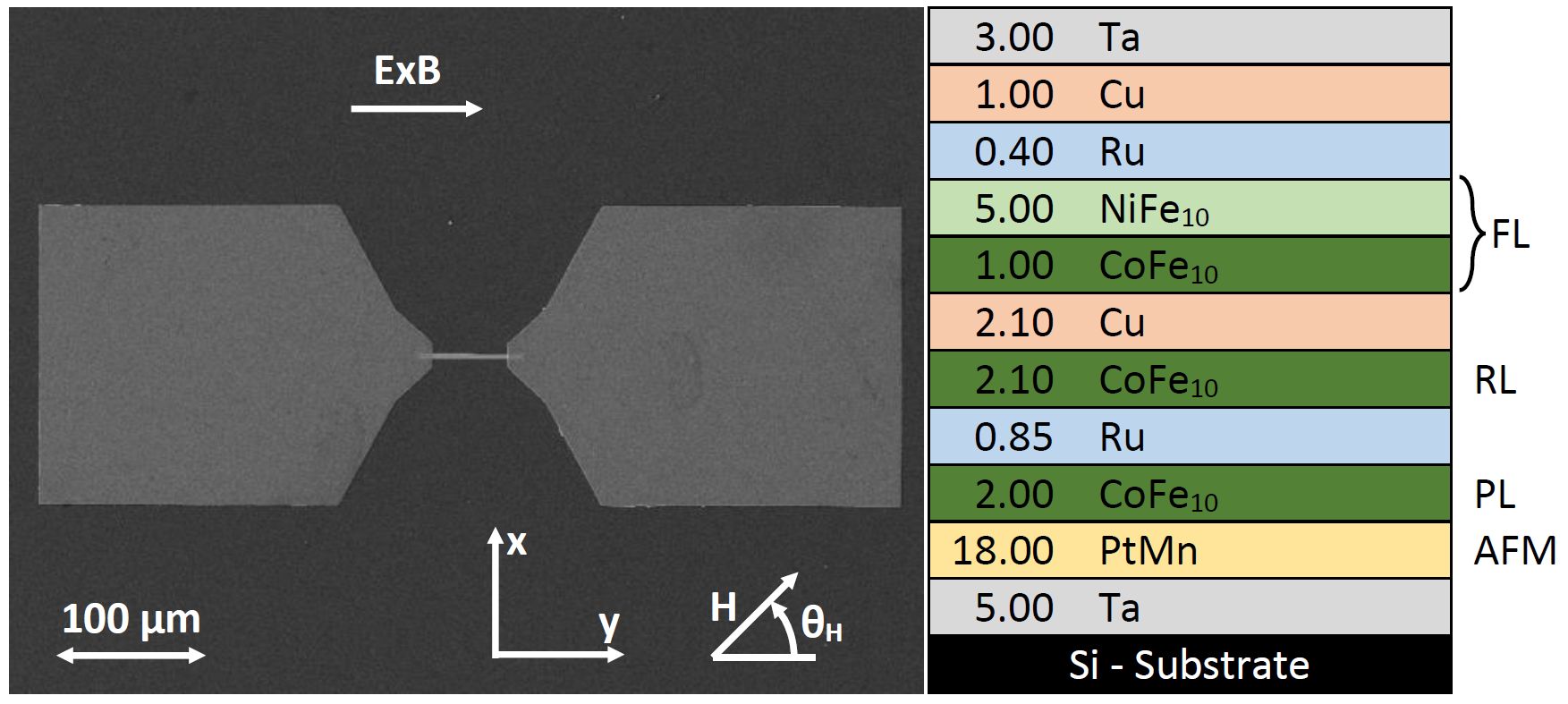}
	\caption{Sample overview and layer structure (layer thickness in nm).}
\label{fig:sample_layers}
	\end{center}
\end{figure}

For test purposes, a spin valve giant magnetoresistance (SV-GMR) sample\cite{zietek2015rectification}, fabricated by means of magnetron sputtering, was used (Fig. \ref{fig:sample_layers}.) It consists of a composite NiFe/CoFe (6 nm) free layer (FL)  and CoFe (2.1 nm) reference layer (RL) separated by a Cu (2.1 nm) spacer. RL is antiferromagnetically coupled through a thin Ru layer with a CoFe (2.0 nm) pinned layer (PL) that is deposited on an antiferromagnetic (AFM) PtMn (18 nm). After the deposition process the GMR stack was annealed in the in-plane magnetic field in order to induce an exchange bias between AFM and PL, which is oriented paralelly to the stripe axis. Next, micro-stripes of $5\times 70$ $\mu m^{2}$ were patterned using direct write laser lithography, ion-etching and lift-off process. A representative microscope image of the fabricated device is presented in Fig. \ref{fig:sample_layers}.

\section{\label{sec:summary}Results and discussion}

\subsection{\label{sec:stepResponse}Step response of the system}
\begin{figure}[h!]
\begin{center}
	\includegraphics[width=\columnwidth]{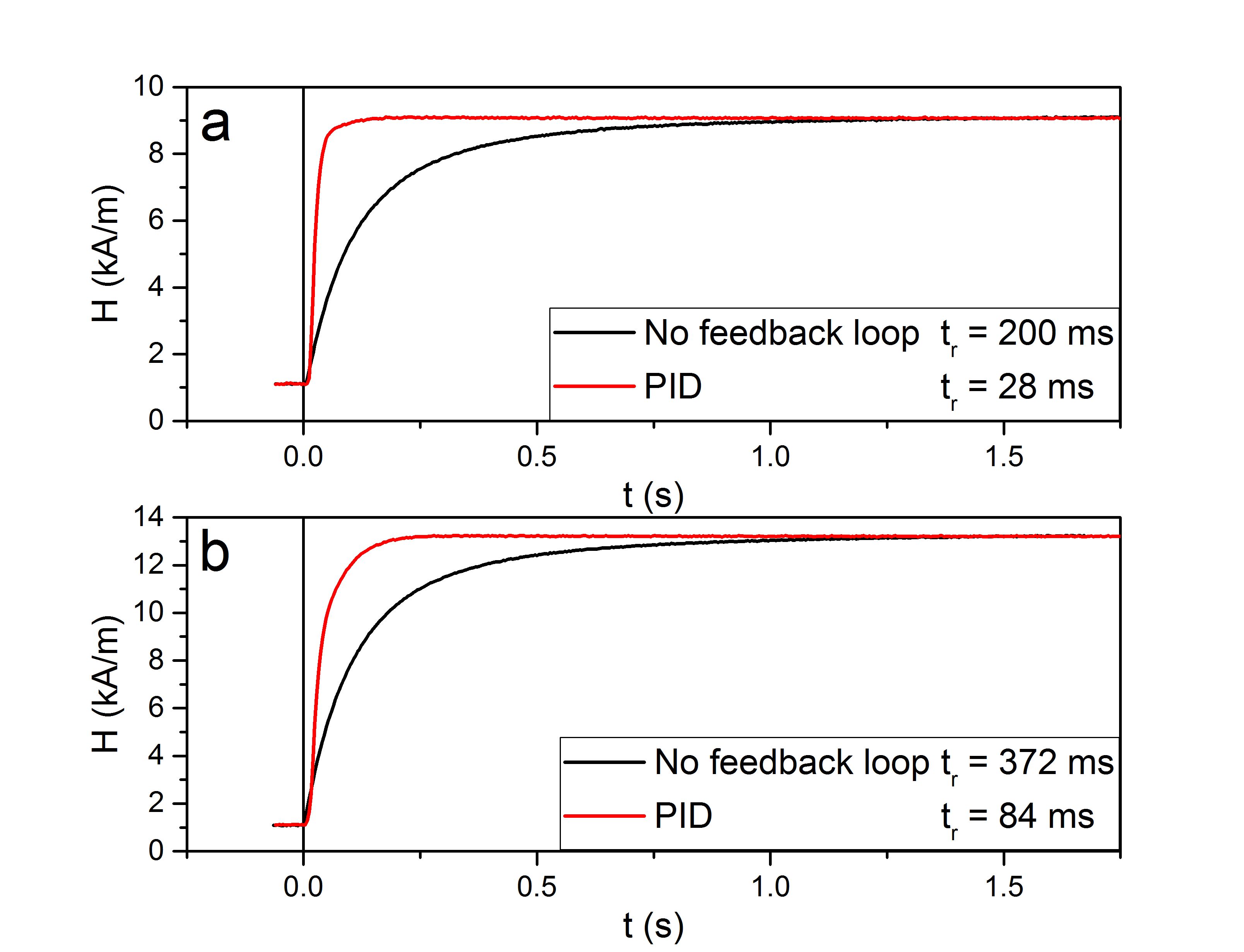}
	\caption{System response comparison for magnetic field H=7.9 kA/m (a) and H=12 kA/m (b) step, $t_r$ are the time constants.}
\label{fig:pid_all}
	\end{center}
\end{figure}

Measurements of magnetic field (H) versus time are presented in Fig. \ref{fig:pid_all}. Black lines represent a response to change in power supply control voltage from 0 V to 0.2 V and 0.3 V, respectively. Red lines represent the results of running a PID controller to obtain the same magnetic field change, as during the application of a voltage step.

The initial value of the magnetic field for all measurements is greater than zero due to remanence of cores, which is impossible to eliminate when using a voltage step, therefore the initial value for the PID controller was set to the corresponding magnetic field obtained using a voltage step.

\begin{figure}[h!]
\begin{center}
	\includegraphics[width=\columnwidth]{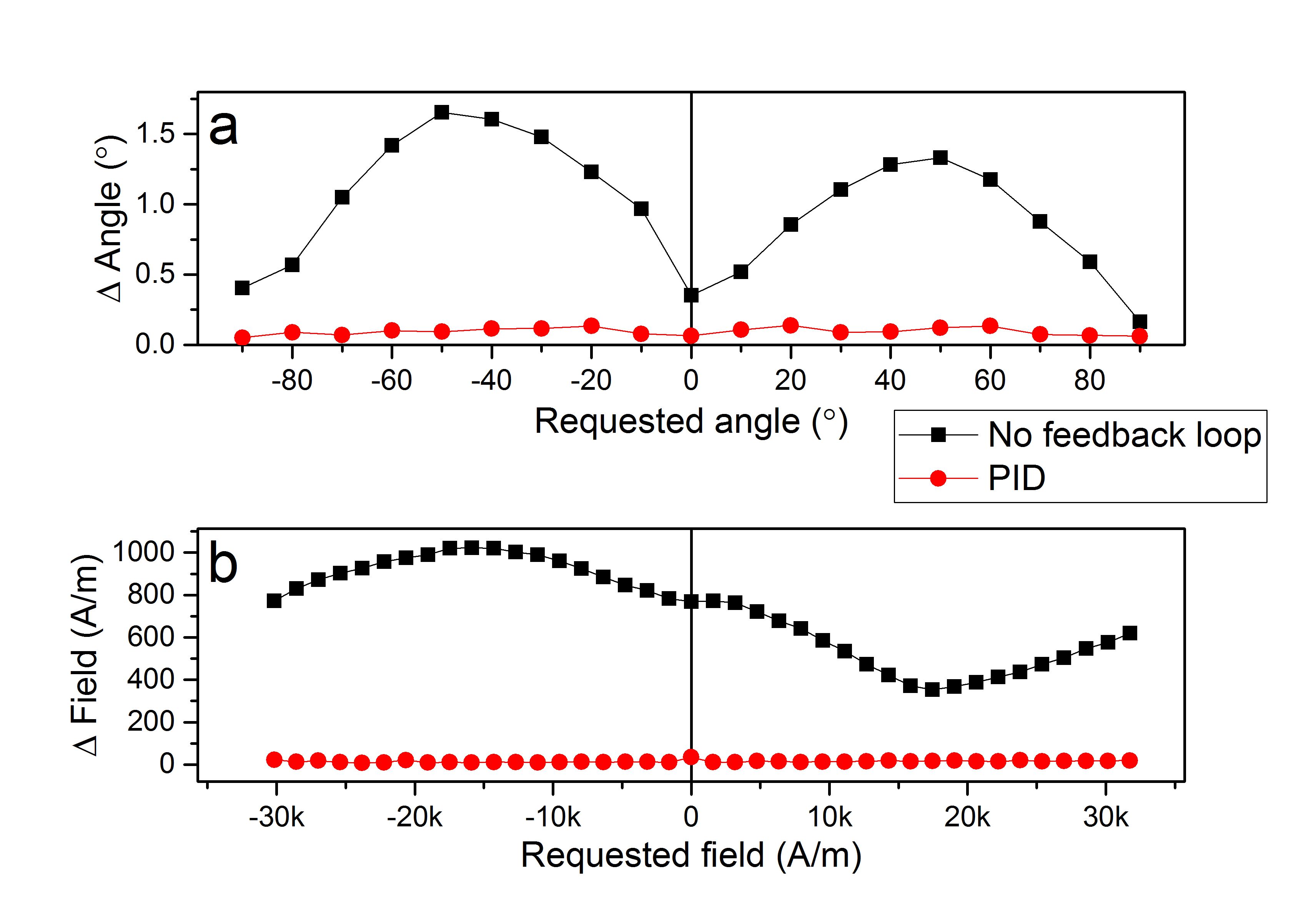}
	\caption{Measurements of the steady state error of angle (a) and magnitude of magnetic field H (b) for cases without a feedback loop and with a PID feedback loop.}
\label{fig:field_err}
	\end{center}
\end{figure}

To verify the strength and angle of the magnetic field vector applied in the center between poles the steady state error (offset field and angle difference between the requested and obtained values) was measured (Fig. \ref{fig:field_err}.) when using the PID controller and without the feedback loop.

As a result a significant improvement of rise time was observed, as well as the ability to cancel the remanence field of cores (steady state error). The maximum field difference for the case with PID is 78 A/m and the angle difference shows a value of $0.13^{\circ}$, compared to 1024 A/m and $1.65^{\circ}$ without the feedback loop.

\begin{figure}[h!]
\begin{center}
	\includegraphics[width=\columnwidth]{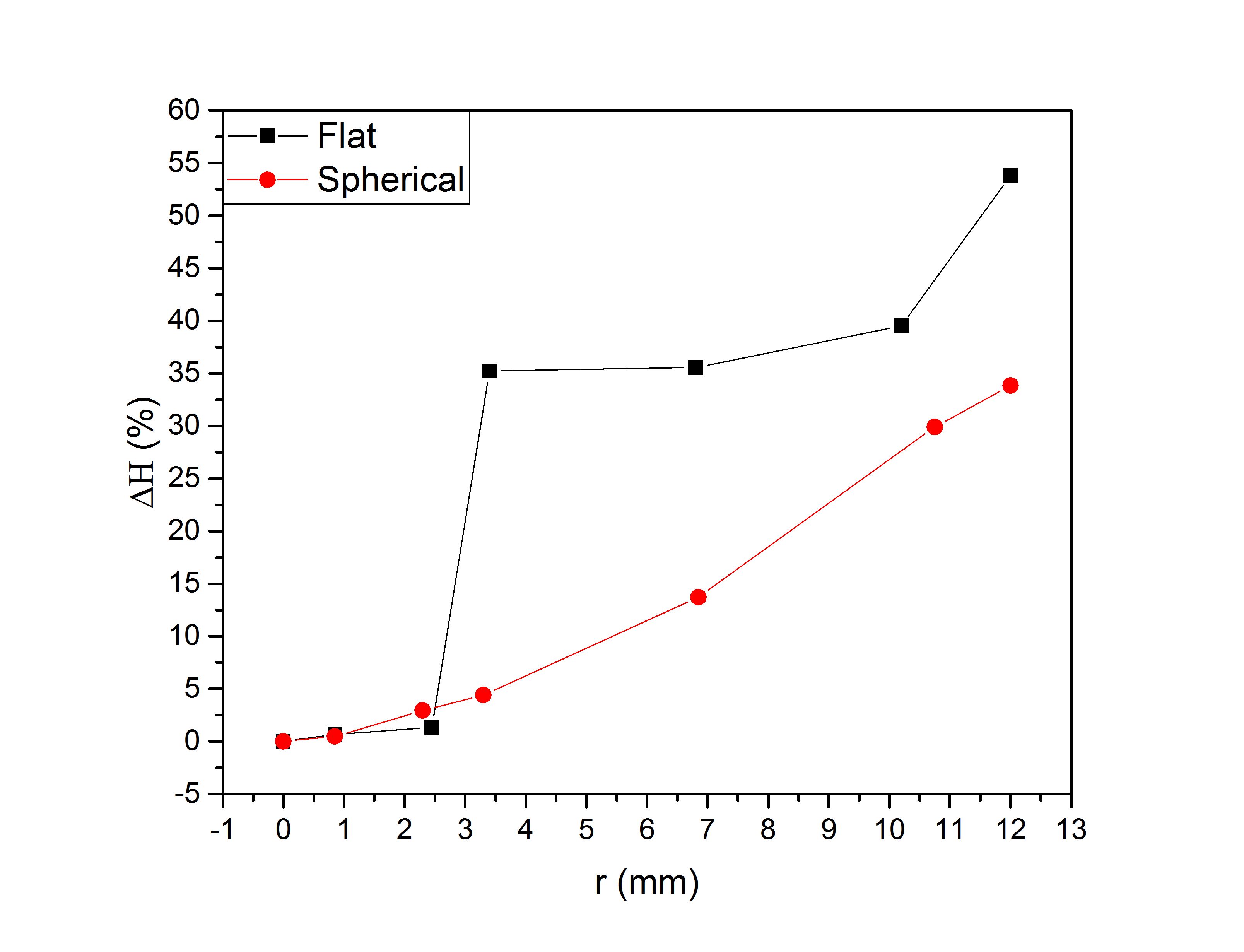}
	\caption{Magnetic field magnitude error for cases with flat and spherical poles.}
\label{fig:poles_H}
	\end{center}
\end{figure}

\begin{figure}[h!]
\begin{center}
	\includegraphics[width=\columnwidth]{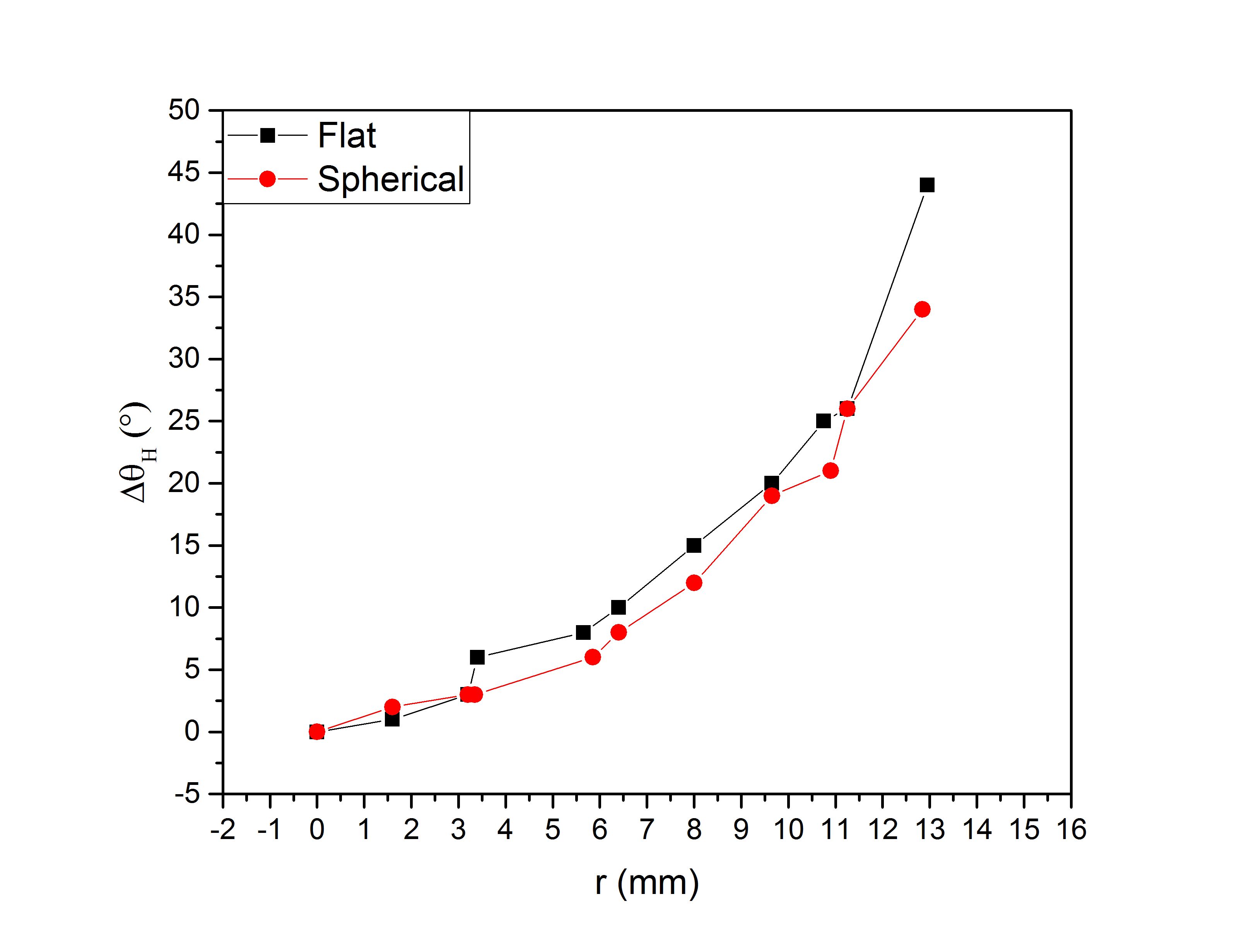}
	\caption{Magnetic field angle error for cases with flat and spherical poles.}
\label{fig:poles_theta}
	\end{center}
\end{figure}

Measurements of the magnetic field vectors in an area between the cores were performed to determine the uniformity of the field. Results of the measurements were analysed as described in Sec. \ref{sec:quadElectromagnet} and are presented on Figs. \ref{fig:poles_H} and \ref{fig:poles_theta}.

\subsection{\label{sec:angularMeasurement}Angular measurements of spin valve GMR}
In this section we show the reliability of the presented experimental setup by performing static (GMR) and
dynamic (SD) measurements of our spintronic device described in details in Sec. \ref{sec:sample}.

\begin{figure}[H]
\centering
\includegraphics[width=\columnwidth,height=20cm,keepaspectratio]{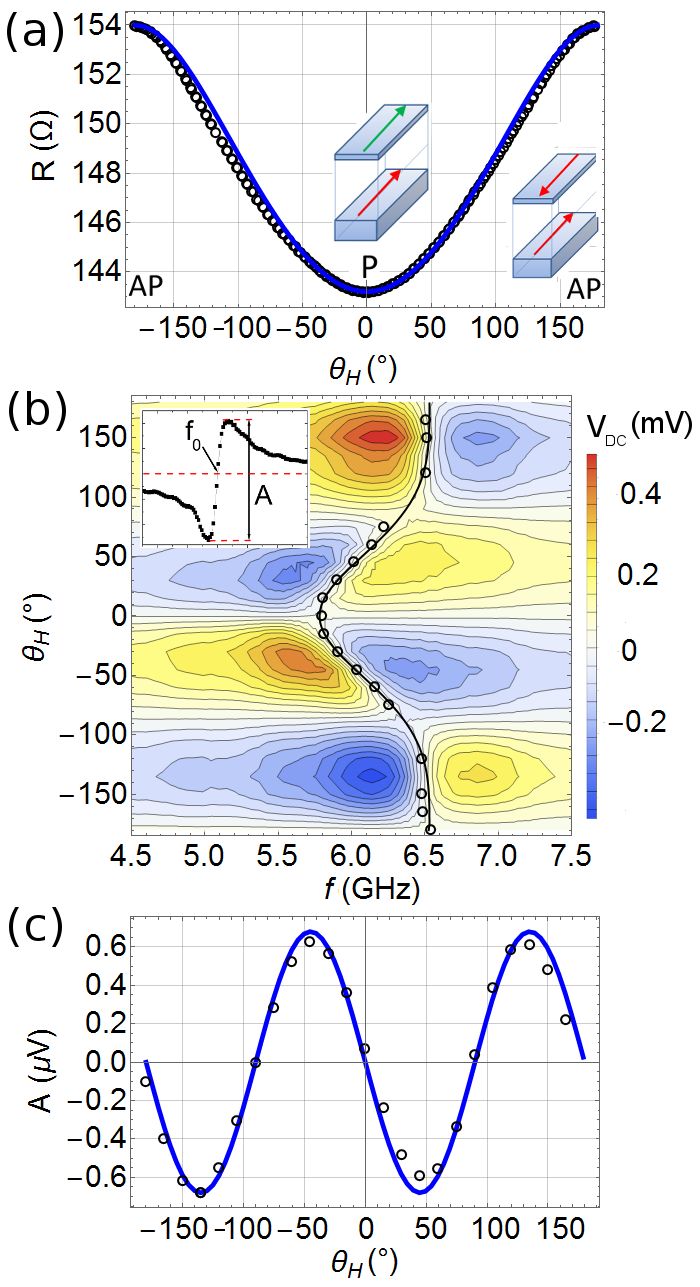}
\caption{(a) Angular dependence of resistance: measured experimentally (black points) and predicted by the macrospin model (blue solid line), P and AP denote the parallel and antiparallel configuration of magnetizations of RL and FL, (b) the two-dimensional map of the $V_{DC}$ signal: examples of experimental resonance frequency ($f_0$) angular dependence (open circles) are compared with the theoretical prediction (black solid lines), (c) angular dependence of the amplitude of the $V_{DC}$ signal: experimental data (black points) and theoretical dependence described by Eq.\ref{eq:katvdc}. Inset: example of the SD antisymmetric spectrum measured at an angle 45$^{\circ}$: $f_0$ denotes resonance frequency, the $A$ is the experimental peak-to-peak amplitude of the $V_{DC}$ signal: the sign of $A$ is negative (positive) when the first extremum of the $V_{DC}$ spectrum is minimum (maximum).}
\label{fig:mapa}
\end{figure}

First we show the angular-dependent resistance measurements. The external magnetic field was set to H=31.8 kA/m.
Next, the direction of the magnetic field was changed from -180 to 180 degrees with respect to the long axis of the GMR microstrip.
The results are shown in Fig.\ref{fig:mapa}(a). The measured resistance R($\theta$) follows the cosine dependence as predicted
for the GMR structures\cite{zietek2015rectification,steren1995angular,barnas1997angular}:
\begin{equation}
R(\theta)=R_P + \frac{\Delta R}{2} (1-\cos\theta)
\label{eq:gmrkat}
\end{equation}
In the above Eq.(\ref{eq:gmrkat}) $\Delta R \equiv R_{AP} - R_P $ is the magnetoresistance, i.e. the difference in a resistance of parallel(P) and antiparallel(AP) states (c.f. Fig.\ref{fig:mapa}(a)).
Next, we measured the SD voltage dependence on external magnetic field direction, and the results are presented in Fig.\ref{fig:mapa}(b). 

In order to verify our experimental results, we adapted the analytical model that was previously developed for an
exchange-biased GMR structure\cite{zietek2015rectification}. Here, the  model accounts for the exchange bias direction parallel to the strip's long axis.

First, we compared the experimental and theoretical dependence of resonance frequency on magnetic field angle. This dependence is visible in Fig.\ref{fig:mapa}(b) as a frequency shift of the experimental $V_{DC}$ spectra for different magnetic field angles. According to the macrospin model (cf. Ref.\onlinecite{zietek2015rectification}), the resonance frequency in the limit of small damping ($\alpha^2 \rightarrow 0$) can be expressed as: 

\begin{equation}
f_0 \approx \frac{1}{2\pi}\frac{\gamma_e}{\mu_0 M_S \sin\theta_M}
\sqrt{   \left(  \frac{\partial ^2 U}{\partial \phi^2}\frac{\partial^2 U}{\partial \theta^2} - \left[ \frac{\partial^2 U}{\partial\phi\partial\theta} \right]^2  \right)   } 
\label{eq:frek}
\end{equation}
where $M_S$ stands for saturation magnetization, $\gamma_e$ is the gyromagnetic ratio, $U(\theta,\phi)$ is the magnetic total energy, and partial derivatives are calcualted at stationary magnetization angles $\theta_M$ and $\phi_M$, which are determined by the local minimum of $U$. Similarly as in Refs.\onlinecite{zietek2015interlayer,zietek2015rectification}, the total magnetic energy $U$ includes the shape anisotropy, the magnetocrystalline anisotropy in the $x$ direction, as well as the interlayer exchange coupling energy term. It was also assumed that the Oersted field has only one component, i.e. $H_{Oe,x}$ perpendicular to the long axis. The calculated angular dependence of the resonance frequency is shown in Fig.\ref{fig:mapa}(b) as a solid line.

As one can see the theoretical prediction agrees well with the experimental result. In both cases the resonance frequency changes by 0.7 GHz while the external field rotates by $180^\circ$. 

Next, we used our model to compare the theoretical and experimental angular dependence of the $V_{DC}$ signal amplitude.
Based on the results presented in Ref.\onlinecite{zietek2015rectification}, we can express the $V_{DC}$ signal as an antisymmetric resonance curve:
\begin{equation}
V_{DC} \approx \frac{A \sin2\theta (f^2-f_0^2)}{(f^2-f_0 ^2)^2 - \sigma^2 f^2} 
\label{eq:katvdc}
\end{equation}
where $f_0$ and $\sigma$ denote the resonance frequency and resonance curve width (FWHM) respectively, $\theta$ is the external magnetic field angle, and $A$ stands for the amplitude of the $V_{DC}$ signal given by:
\begin{equation}
A \approx  - \eta \frac{\gamma_e }{2} \pi I H_{Oe} \Delta R  \left( H_K+H+ M_S (N_z - N_x) \right)
\label{eq:amp}
\end{equation}
In the above Eq.(\ref{eq:amp}) $I$ and  $H_{Oe}$  are the amplitudes of the alternating current and associated Oersted field, respectively, while $N_{z(x)}$ denotes the demagnetizing factor in the $z(x)$ direction. The phenomenological factor $\eta$ was introduced in order to account for all possible losses of microwave signal in the experimental setup\cite{zietek2015interlayer}. Similarly as in the case of resonance frequency calculations, we assumed the following values of magnetic parameters: saturation magnetization $\mu_0 M_S = 1.06$ $T$, uniaxial anisotropy field $H_k = 2.8$ $kA/m $, demagnetizing factors: $N_z = 0.997311, N_x = 0.00246728$ and $N_y=1-(N_z+N_x)$. 
\\
 According to the  Eq.(\ref{eq:katvdc}), the  spin-diode signal follows $\sin 2 \theta$ dependence, what is also visible in the case of experimental data shown in Fig.\ref{fig:mapa}(c). 

\section{\label{sec:summary_final}Summary}
The experimental setup for static and dynamic angular magnetic measurements has been developed and presented. 
As a result of application of the quadrupole electromagnet with a dedicated feedback control system, we have been able to  
generate an arbitrary in-plane magnetic field vector. Moreover, in order to increase the area with the uniform magnetic field vector, we optimized the shape of the electromagnet cores. Therefore we have eliminated a possible magnetic measurement inaccuracy that may occur in a standard sample-rotating setup. The experimental system was tested by measuring the angular dependence of the resistance and spin diode effect in the spin-valve GMR micro-stripe. Both static (resistance) and dynamic (spin-diode effect) angular dependencies were compared with the theoretical predictions that confirmed the reliability of the presented measurement setup. 

\section*{Acknowledgement}
We acknowledge the Polish National Center for Research and Development grant No. LIDER/467/L-6/14/NCBR/2015.

P.O. acknowledges Dekaban Fund at the University of Michigan for the financial support.

\bibliography{bibliography}

\end{document}